\documentclass[a4paper,preprint,unsortedaddress,onecolumn]{revtex4-1}

\usepackage[fleqn]{amsmath}
\usepackage{amssymb}
\usepackage[dvips]{graphicx}
\usepackage{color}
\usepackage{tabularx}
\usepackage{algorithm}
\usepackage{algorithmic}


\begin{document}

\title{Grand Canonical Adaptive Resolution Simulation for Molecules with Electrons: A Theoretical Framework based on Physical Consistency}
\author{Luigi Delle Site}
\affiliation{Institute for Mathematics, Freie Universit\"at Berlin, Germany}
\email{luigi.dellesite@fu-berlin.de}
\begin{abstract}
A theoretical scheme for the treatment of an open molecular system with electrons and nuclei is proposed. The idea is based on the Grand Canonical description of a quantum region embedded in a classical reservoir of molecules. Electronic properties of the quantum region are calculated at constant electronic chemical potential equal to that of the corresponding (large) bulk system treated at full quantum level. Instead, the exchange of molecules between the quantum region and the classical environment occurs at the chemical potential of the macroscopic thermodynamic conditions. T he Grand Canonical Adaptive Resolution Scheme is proposed for the treatment of the classical environment; such an approach can treat the exchange of molecules according to first principles of statistical mechanics and thermodynamic. The overall scheme is build on the basis of physical consistency, with the corresponding definition of numerical criteria of control of the approximations implied by the coupling. Given the wide range of expertise required, this work has the intention of providing guiding principles for the construction of a well founded computational protocol for actual multiscale simulations from the electronic to the mesoscopic scale.\\

\end{abstract}
\maketitle
\section{Introduction}
Multiscale modeling in molecular simulation has been a very present theme in
research during the last two decades. The idea of coupling different scales
within a unitary framework led to different lines of research in different
fields and disciplines \cite{entropy}. One major technical and conceptual
challenge is the coupling of the quantum/electronic scale with the classical
statistical/thermodynamic scale. For example, the solvation of a molecule in
water whose details of the making and breaking of hydrogen bonds play a key
role in the hydration properties, needs a quantum description close to the
solute while a classical description of the bulk, which acts as a reservoir of
energy and particles for the quantum region, is sufficient. The conceptual relevant point is
that in order to have proper physical consistency in the quantum region one
should be able to reproduce quantum/electronic properties as close as possible
to those of an equivalent subsystem of a large system treated at full quantum
level. From a rigorous pint of view, due to electronic correlations, the electronic spectrum of eigenvalues of a subsystem
embedded in a larger (quantum) system will always be different from the
spectrum of a subsystem embedded in a classical system, however the
equivalence of properties can be considered at statistical mechanics level
rather than at single-electron wavefunction level. In fact if we consider the
quantum subsystem as a Grand Canonical (open) ensemble which exchanges
particles with a reservoir, then we are flexible in choosing the specific
characteristics of the reservoir as long as particles (nuclei and electrons)
are exchanged according to a Grand Canonical statistics (see corresponding discussion in Ref.\cite{physrep}). In this way a quantum subsystem  coupled to a classical environment and having the same electronic chemical potential of a large quantum system will be statistically equivalent, at
electronic level, to a quantum subsystem embedded in a large quantum system. A
rigorous treatment of a quantum subsystem embedded in a larger environment at
arbitrary coarser resolution was considered in Ref.\cite{emch}; by employing
the projector operator technique the equations for the subsystem and the
coupling to the rest of the system were derived. The solution shows that the
Grand Canonical density matrix describes the properties of the subsystem. The coupling to the environment occurs through an explicit interaction potential
between the particles of the subsystem and those of the reservoir, plus a term function of particles'
creation ad annihilation operators. As in any Grand Canonical ensemble
derivation, the implicit hypothesis is that the coupling potential corresponds
to an energy which is negligible compared to the energy of the quantum subsystem (see
e.g. \cite{huangbook}). The derivation of Ref.\cite{emch}, expresses in formal
terms the physically intuitive description given before, that is considering
the quantum subsystem as a Grand Canonical ensemble of electrons and
nuclei. This is the conceptual starting point from which I will build the theoretical
framework for a computational protocol that embeds a Grand Canonical
quantum systems in a reservoir of classical molecules. The idea proposed here is different from current, so called, open
boundaries QM/MM; in fact the coupling proposed here assumes strict physical
consistency in the form of necessary conditions. For physical consistency is
meant that for every approximation, strict, conceptual and numerical, {\it a priori}, criteria of control
of validity are defined. This implies that the level of violation of the chosen 
physical hypothesis
automatically defines the level of accuracy of a calculation. For the
description of the classical
environment I propose the Grand Canonical Adaptive Resolution (GC-AdRess)
scheme \cite{prx,njp,physrep}. Such a scheme has been shown to assure that a molecular liquid treated
at different concurrent resolutions exchanges molecules between different regions
at the appropriate chemical potential defined by the macroscopic thermodynamic
conditions desired \cite{jcpsimon,jctchan}. Within the context of embedding a QM region in a classical environment (reservoir) GC-AdResS can assure that the exchange of molecules (classical nuclei) between the quantum region and the classical reservoir occurs at the proper macroscopic (molecular) chemical potential. Moreover, GC-AdResS allows to describe
the largest portion of the classical environment at a very simple
coarse-grained level and thus it is computationally convenient. 
As anticipated above, the number of electrons in the
quantum region is regulated by electronic structure calculations, for a given
nuclei configuration, done at fixed
electronic chemical potential, as if, from the electronic point of view, the quantum system is virtually embedded in a larger electronic system (full quantum system of reference).   
For completeness in the next section I will report the basics of QM/MM methods with open boundaries and underline its practical efficiency and its conceptual pitfalls. Next the approach of electronic structure calculations for open systems at given chemical potential is reported, followed by a description of the essential aspects of GC-AdResS needed here. Finally the scheme of dynamic coupling of
GC-AdResS with the QM region is shown. A section regarding critical aspects of
the idea and possible practical solution, together with some final conclusions
closes the paper. Finally, it must be underlined that the focus of this paper
is not the numerical implementation and its corresponding efficiency, rather this paper represents a proposal of a theoretical framework with physical consistency which can certainly lead to a numerical implementation. The research communities
involved are technically far from each other and the intention of the
paper  is to bridge this gap, stimulate cooperation on the basis of a
common ground provided by the theoretical framework proposed here and possibly finalize it in a working code. 

\section{QM/MM with open boundaries}   
Concurrent, adaptive molecular resolution simulations with electronic degrees of freedom have been formulated so far as an an extension of the QM/MM method \cite{qmmm0,qmmm01}; in such an extension an open boundary between the QM and the MM is (effectively) mimicked \cite{qmmm1,qmmm2,qmmm3,qmmm4}. This allows to estimate the effect of a change of the number of molecules,  in the QM region, due to its interaction with the bulk. Such an approach, as said before, mimics a dynamical system with open boundaries, but as a matter of fact, the change of the physics in the QM region does not happen in a dynamic way while, at the same time, thermodynamic consistency between the hypothetical bulk and the quantum subregion also comes in question. For thermodynamic consistency it is meant that the QM subregion of the QM/MM system must have the same macroscopic and microscopic (electronic) statistical mechanics/thermodynamics properties of an equivalent subregion of a large full quantum simulation.
 The original QM/MM method is based on the hypothesis that the electronic
 properties of the large environment around a certain region are not relevant,
 thus the treatment of its molecules can be done at classical level, i.e., the
 environment plays the role of an effective particle-based classical
 thermodynamic bath with fixed number of molecules. From the technical and the conceptual point of view, the
 key aspect of the QM/MM method is the coupling between the two regions. The
 original scheme considers the QM and the MM region fixed regarding the number
 of particles in each region (i.e. there is no particle exchange). The
 progress in time in engineering the interface/coupling setup has brought
 QM/MM methods of the last generation into the category of methods which can
 technically mimic open boundary systems with a variable number of molecules. It
 must be clarified that the explicit intention of the developers is of
 technical character, since the main aim of the method is that of minimizing
 the computational errors in the QM part due to the interface with the MM
 part in the standard QM/MM. In this sense, the adaptive QM/MM cannot be considered a systematic
 attempt to build a procedure with strict physical consistency for an open
 boundary quantum system. The major problem in such schemes is that the {\it
   ``fluctuations''} of number of molecules in the QM region implies a drastic
 change of the total energy of the system. In order to minimize this problem a
 buffer or transition region is defined. A paradox then emerges, that is a
 molecule in the hybrid region would have a nonphysical fractional quantum character. In current QM/MM methods, from the technical
 point of view one needs that either molecules are treated at full quantum
 level or (as in the standard QM/MM scheme with rigid boundaries) as classical
 molecules. This problem is solved with a principle common to most of the
 current adaptive QM/MM methods, that is at each time-step of the simulation,
 the buffer region is partitioned in different subsets. Next, the (standard)
 QM/MM potential is defined for all the possible partitionings and, for each
 partitioning, the molecules of the corresponding subset of the buffer are
 included in the QM region. Thus, for each of these ``extended'' QM regions a
 standard QM/MM calculation is done. The total potential is then defined as a
 weighted average of these individual potentials: $U({\bf
   r})=\sum_{i}^{M}f_{i}({\bf r})U_{i}({\bf r})$, where each $U_{i}({\bf r})$
 corresponds to one of the $M$ partitioning of the system in a group of QM
 molecules and a group of MM molecules and $f_{i}({\bf r})$ is the weighting
 function. The weighting function $f_{i}({\bf r})$ is function of the
 coordinates of the molecules and can be expressed in terms of single molecule
 switching functions. The switching function is constructed to follow the
 general principle that the quantum energy of molecules far way from the
 active site counts less that the energy of those which are closer to the active site. Next, the
 dynamical evolution, based on such potential energy, is performed and it
 creates a new configuration on which the partitioning step is applied once
 again and the procedure for the calculation of the potential energy follows
 as described before. Such an idea have been successfully implemented by
 several groups, with specific choices of the switching function, of the
 partitioning scheme  or based on a force interpolation rather than an energy
 interpolation \cite{qmmm1,qmmm2,qmmm3,qmmm4}. Despite numerical results are
 (at least within a basic accuracy) encouraging, from the technical point of
 view, due to the partitioning procedure, the number of QM calculations
 required (compared to standard QM/MM studies) may represent an expensive
 effort. Most importantly, besides the computational effort, at conceptual level the theoretical
 framework as open boundary approach is not yet solid and actually may easily
 lead to artificial results. As underlined before, the process of variation of
 number of molecules in the QM region is achieved via an artificial path, in
 the sense that the variation of molecules in the QM region does not follow
 or satisfy rigorous principles of statistical mechanics. For example, it has
 not be shown that the variation of number of nuclei and electrons in
 the QM region occurs according to the proper chemical potential for the nuclei and the proper chemical potential for the electrons according to the thermodynamic conditions expected; the fulfillment of such conditions it is mandatory for claiming physical and chemical consistency. Hence, the thermodynamic
 conditions, in which the QM region effectively is, may not correspond to the
 conditions expected and thus results may be artificial. In turn, this implies that one should always check, case
 by case, that the adaptive QM/MM study reproduces some reference results
 (from experiment or larger QM/MM calculations), as it is actually done nowadays (see discussion in Ref.\cite{revqmmm}). An exception is represented by the approach followed by Heyden and Truhlar whose QM/MM scheme \cite{ht1} may be recast into a unified Lagrangian approach which implicitly provides physical consistency; despite possible technical difficulties this idea is promising and is part of an ongoing project \cite{ht2}. Instead, as underlined before, the
 idea reported in this paper is based on the thermodynamic and
 statistical mechanics consistency; nuclei and electrons are introduced in the QM region (or removed from the QM region) though a 
 ``dynamical insertion/removal''. This process is in turn realized via a molecular adaptive resolution technique. Moreover, due to
 physical rigorous conditions imposed at the various interfaces, the MM part
 can be further simplified by the use of classical coarse-grained molecular
 representation of the large bulk. In this perspective, the first question to address is how to treat at
 conceptual and computational level an open system of electrons, this aspect is discussed in the next section.
\subsection{Electronic systems in a Grand Canonical ensemble: Density Functional calculations  at fixed chemical potential}
Traditional constant-$N$-Density Functional Calculations (DFT) are by now known
across the entire molecular simulation community.
Within the Born-Oppenheimer approximation nuclei can be considered classical
objects and for a fixed nuclei configuration the corresponding electronic
density of ground state is calculated through the minimization of the energy functional,
according to the Hohenberg-Kohn theorem \cite{kohn}. Next the Hellmann-Feynman forces on the nuclei can be calculated and the nuclei moved according to classical equations of motion to a new configuration (see e.g.\cite{bookdft}). This is
in brief the essence of modern {\it ab initio} Molecular Dynamics (AIMD).
However, keeping in mind the discussion of the previous sections, the standard
DFT approach when the quantum system is coupled to another system through open
boundaries, comes into question. The solution used in open boundaries QM/MM is
artificial and it is likely to violate (in an uncontrolled way) mandatory criteria of physical
consistency; however a solution, with physical consistency at statistical
mechanics level, as suggested before, is to consider the electronic system as a
Grand Canonical ensemble. The Theoretical
foundations of DFT for open systems has been established for long time along the
lines suggested by the Capitani, Nalewajski, Parr (CNP) theory \cite{cnp} (see
also \cite{bookdft}). At constant number of electrons, $N$, one has:
$\delta(E[\rho_{e}]-\mu_{e}[\int \rho_{e}({\bf r})d{\bf r}-N])=0$, where $\rho_{e}$ is the three-dimensional electron density, $\mu_{e}$
    is the Lagrange multiplier associated to the condition of normalization of
    $\rho_{e}$ to $N$; for an open system at constant $\mu_{e}$, $N$ is the corresponding
    Lagrange multiplier. A rigorous derivation of the idea above implies the
    introduction of an electronic equivalent of the Grand Potential:
    $Q[\rho_{e}]=E[\rho_{e}]+N\mu_{e}$ and an associated minimization:
    $min_{N}(E[N,\rho_{e}])+\mu_{e}N$ \cite{newa}. Various techniques to perform such a
    calculations have been developed and applied in computational schemes
    \cite{bureau,ali,marzari,auer}. In essence the relevant distinction between the
    Canonical and Grand Canonical ensembles for an electronic system is that
    in the Canonical ensemble, it is assumed a fixed (constant) number of
    electrons for each microscopic configuration taken by the system. In such
    a case the chemical potential is an average over all the
    configurations. In the GC ensemble instead the chemical potential is fixed
    for each microscopic configuration; this implies that the number of
    electrons is variable and the fixed $N$ of the Canonical ensemble is now
    substituted by the average number of electrons. One point to underline is
    that $N$ may not be an integer anymore, the physical legitimization of
    such a point has been discussed in a convincing manner \cite{ref1,ref2}.
In our case it will have some practical consequences for the coupling to the
classical system which will be addressed later on.
At this point, we need to consider one further aspect, that is the practical definition of proper electronic chemical potential (electronic chemical potential of
reference); such a chemical potential is defined as the electronic chemical potential of the large bulk that ideally acts as a reservoir/environment.
In practice, the reference chemical potential should be be calculated, in a
separate calculation, using a standard constant-$N$ DFT approach for an
auxiliary system that reasonably represents a generic large bulk.  For example, let us suppose that one wants to simulate biomolecules in water; bulk water is the environment of the
solvation region, thus the electronic chemical potential corresponds to that of bulk water. As a consequence, the electronic chemical potential calculated for a reasonable large bulk water system can be used as the electronic
chemical potential of reference for different solvation problems involving
liquid water. This implies that its calculation is required only once and can be used for all the cases/situations
where the reservoir has the same characteristics. Finally, an important point regards the coupling
with classical environment . This coupling enters into the electronic Hamiltonian,
the Coulomb interaction with the environment, which acts as
an external potential reads: $H_{ext}=\sum_{i=1,W}\int \frac{\rho_{e}({\bf
    r})q_{i}}{|{\bf r}-{\bf
    x}_{i}|}+\sum_{I=1,M}\sum_{i=1,W}\frac{Z_{I}q_{i}}{|{\bf R}_{I}-{\bf
    x}_{i}|}$, where $q_{i}$ is the charge of the $i$-th atom in the classical
region, ${\bf x}_{i}$ its position, $W$ is the total number of atoms
considered in the classical region. In this regard a crucial control criterion
of physical consistency (hypothesis under which one has a Grand Canonical
ensemble for the QM region) is that $\langle H_{ext}\rangle << \langle H_{QM}
\rangle$ at any time step of the simulation (i.e. for any configuration in space of nuclei and classical environment). A similar condition has been already used in the simulation of
classical systems (see \cite{njp}). It must be underlined that also in QM/MM
simulations this condition should be always fulfilled, in fact if the inequality does not
hold it means that the coupling with the environment is dominant and it
determines the physics of the system. Thus, it does not make sense to talk
about quantum properties of the subsystem in a quantitative manner since it is
very likely that they are entirely artificial (at electronic level) or at the
best vaguely qualitative. {The specific form of the external term of interaction in the QM Hamiltonian (Coulomb or Lennard-Jones type), which is a major problem in QM/MM \cite{tru}, is, from the conceptual point of view, not relevant in our treatment. However I have assumed and will assume in the rest of the paper that the interaction is always of electrostatic type. In any case, it must be underlined that any Hamiltonian used in the QM/MM approach can be actually used also in this scheme.} At this point one has the physical principles to
describe a quantum subsystems and calculate electronic properties, the
question left is how to build a larger thermodynamic environment, dynamically
coupled to the quantum region, which assures proper exchange of molecules with
the QM subsystem according to the macroscopic thermodynamics of reference. The
coupling with a classical environment that acts as a reservoir of molecules
proposed here is based on the Grand Canonical Adaptive Resolution (GC-AdResS)
method which couples regions of space where molecules have different level of
(classical) resolution with thermodynamic and statistical mechanics
consistency. 
In the next section I describe the basic characteristics of GC-AdResS and then
explain how it can be coupled to the quantum region and assure thermodynamic
consistency to the system.
\section{GC-AdResS: Basics}
\label{adress}
\begin{figure}
   \centering
   \includegraphics[width=0.75\textwidth]{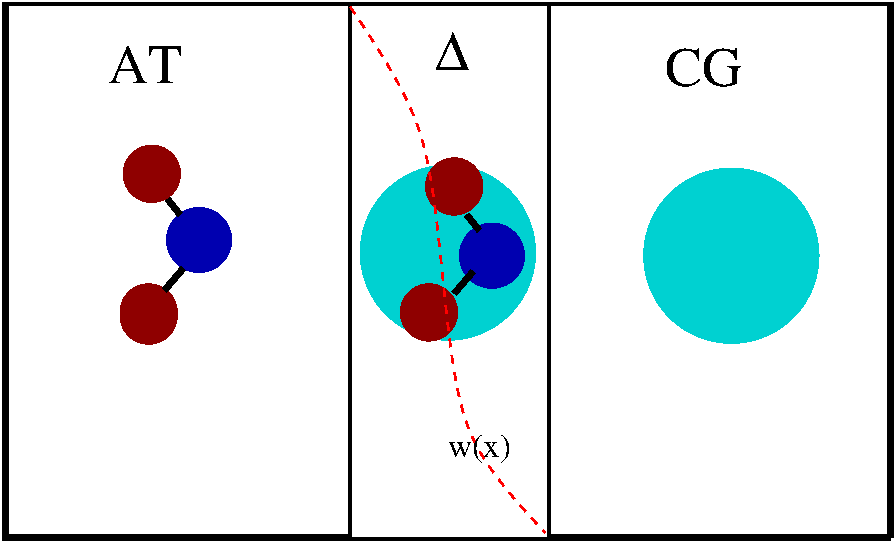}
   \caption{Schematic representation of the AdResS partitioning of the simulation box. AT is the atomistic region where the molecular representation follows the standard route of classical force-fields; $\Delta$ is the transition region where atomistic and coarse-grained forces are spatially interpolated through $x(x)$ to allow for a transition from one representation to another. Additionally a thermodynamic force is applied to the center of mass of the molecules in order to have thermodynamic consistency (i.e. level off the chemical potential of the atomistic and of the coarse-grained representation. Finally, in the CG region the molecules interact only through a coarse-grained potential acting between the centers of mass of pair of molecules.}
 \label{cartoon-adress}
 \end{figure}

The Adaptive Resolution method (AdRess)for Molecular Dynamics \cite{jcp1,pre1} in its original formulation followed the intuitive principle that regions where molecules have different molecular resolution can be coupled in such a way that going from the dynamics of one resolution (region in space) to another should be smooth enough so that the perturbation to the local dynamics of each region is negligible. Following this principle, the computational algorithm was build through a space dependent interpolation for the force between two molecules, $\alpha,\beta$ (see Fig.\ref{cartoon-adress}):
\begin{equation}
F_{\alpha \beta} = w(X_{\alpha})w(X_{\alpha})F_{\alpha\beta}^{AT} + [1 - w(X_{\alpha})w(X_{\alpha})]F_{\alpha\beta}^{CG}
\end{equation}  
where $F_{\alpha\beta}^{AT}$ is the atomistic force and $F_{\alpha\beta}^{CG}$ is the coarse-grained force. The interpolating function chosen is:
\begin{equation*}
    w(x) = \begin{cases}
               1               & x < d_{AT} \\
               cos^{2}\left[\frac{\pi}{2(d_{\Delta})}(x-d_{AT})\right]   & d_{AT} < x < d_{AT}+d_{\Delta}\\
               0 & d_{AT} + d_{\Delta}< x
           \end{cases}
\end{equation*}
where, $d_{AT}$ expresses the extension of the atomistic region, while and $d_{\Delta}$ is the size of the hybrid or transition region; $x$ is the 
$x$-coordinate of the center of mass of the molecule.
$w(x)$ smoothly goes from 0 to 1 in $\Delta$ so that a coarse-grained molecule, as it moves towards the atomistic region, slowly but systematically starts to feel more atomistic environment (interactions) and it becomes an atomistic molecule and vice versa for an atomistic molecule moving towards the coarse-grained region. The system is embedded in a thermostat that supplies or removes the required energy and assure the desired thermodynamic equilibrium. Despite its conceptual simplicity, this approach produced highly satisfactory results for a large variety of systems and physical situations \cite{pol,wat1,wat2,jcpcov,prlado,pccpado,jctckb,jcppara}. However, exactly because it turned to be a numerically satisfactory approach, a deeper conceptual justification of the idea was mandatory. A first step forward involved the analysis of basic thermodynamic relations which should express the coupling of resolution in a more precise way. Such an analysis led to the derivation of a force, ${\bf F}_{thf}(x)$ (thermodynamic force), acting on the center of mass of the molecule in $\Delta$, which assures that the effective chemical potential of the whole system corresponds to that of the atomistic resolution \cite{jcpsimon}. In later developments, ${\bf F}_{thf}(x)$, was derived within a more rigorous framework: the atomistic and the coarse-grained region are now considered as two open regions interfaced by a filter, $\Delta$, and ${\bf F}_{thf}(x)$ is defined through the balance of the Grand Potential. Essentially the open coarse-grained region corresponds to a reservoir for the open atomistic region and vice versa. The explicit condition reads :$p_{AT}+\rho_{0}\int_{\Delta}{\bf F}_{thf}({\bf r})d{\bf r}=p_{CG}$,
  with $p_{AT}$  the pressure of reference of the atomistic system (region), $p_{CG}$ the pressure of the coarse-grained model and $\rho_{0}$ is the molecular density of reference of the atomistic system. A numerical convenient way to calculate the thermodynamic force is to express it as the gradient of the particle density in an iterative form \cite{prl12}: $F_{k+1}^{thf}(x)=F_{k}^{thf}(x) - \frac{M_{\alpha}}{[\rho_{ref}]^2\kappa}\nabla\rho_{k}(x)$,
$M_{\alpha}$ is the mass of the molecule, $\kappa$ a (conveniently) tunable constant, $\rho_{k}(x)$ is the molecular density  as a function of the position in $\Delta$, at the $k$-th iteration.
The convergence criterion depends on the accuracy required for the simulation
but a rule of thumb, $\rho_{final}-\rho_{0}$ should be below $10\%$ in
$\Delta$. The thermodynamic force is calculated in a pre-production run as equilibration of the system and then used for the production run. Moreover the thermodynamic force of the same system at given thermodynamic conditions can be used also in other (separate) simulations; for example if we use water as a solvent of a molecule, the thermodynamic force of water since it is applied in the bulk, is independent of the solvated molecule. Finally, in the last few years it became possible to lie a
mathematical and physical rigorous formalization (with quantitative estimate
of the errors due to the numerical approximations) either in terms of a
global Hamiltonian (H-AdResS) \cite{raff1,raff2} or in terms of Grand Ensemble
approach where the coarse-grained region is a reservoir of energy and
particles for the atomistic region \cite{jctchan,prx,njp,pre16}; needless to
say that both approaches are originated from the same roots and are thus
compatible \cite{kreisepl,ralfepjst,prejinglong}. {Of particular importance, for its coupling to continuum and hydrodynamics and thus for the extension to the scale even beyond the mesoscopic one, is the version of AdResS that goes under the name of Open Boundary Molecular Dynamics (OBMD) (see e.g. \cite{epjstmat})}. In the meanwhile the AdResS
method in any current formulation has been successfully applied to a rather
large variety of liquids in different thermodynamic conditions
\cite{kk15,kk161,kk162,mat1,mat2,mat3,luimu,lujcppi,lucpc,lupolj}. It is important to underline that the coupling scheme allows to interface not only atomistic and coarse-grained resolution but any (classical) molecular representation (i.e. two different atomistic water models, atomistic and path integral representation of molecules \cite{lujcppi,cpcanim,jctcraff}. Here we go beyond the classical representation and the idea of
coupling GC-AdResS to the QM region is that a region of classical atomistic
molecules at high resolution (flexible bonds and angles) acts as interface
between the QM region and a large coarse-grained region. In GC-AdResS routine calculations the thermodynamic
force acting in the region between the classical atomistic molecules and the
coarse-grained region assures thermodynamics consistency between the two
classical models; however we now need also thermodynamic/statistical mechanics consistency between the
QM region and the atomistic region. This means that the exchange of classical nuclei of
the QM region (which enter into the atomistic region and become classical molecules)
and the atomistic molecules (which enter into the QM region becoming classical
nuclei) must occur at the same thermodynamic conditions as in the rest of the
(classical) system. For such an aim, as for $\Delta$ in the MM/CG interface the thermodynamic force is applied in the MM region so that the
molecular density fluctuations in the QM region are 
automatically regulated as in the rest of the system. In fact in the QM region the nuclei are classical objects so that from the classical point of view one can visualize the skeleton of a molecule formed by the nuclei as a classical molecule with a bare nuclear charges. Then the thermodynamic force applied in the MM region couples two different molecular representations (skeleton formed by the nuclei of the quantum molecule and the classical atomistic molecule) with thermodynamic consistency as it has been put forward in Ref.\cite{jcpsimon}. In the next section the details of the coupling are discussed.

\section{QM region embedded in GC-AdResS}
\begin{figure}[h!]
  \centering
  \includegraphics[width=1.1\textwidth]{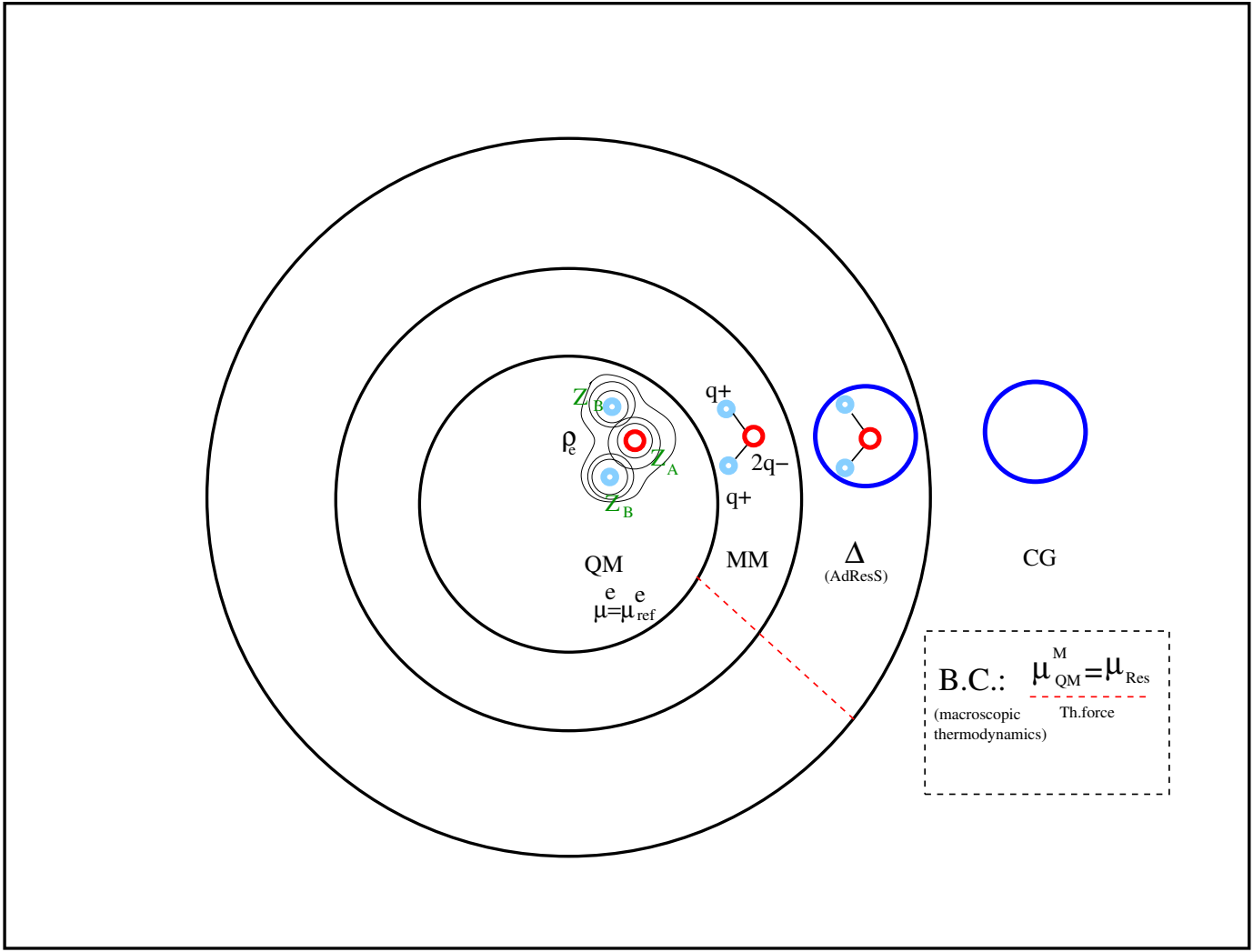}
  \caption{Pictorial representation of the quantum GC-AdResS scheme. ${\mu}_{QM}^{M}$ is the macroscopic chemical potential obtained through the thermodynamic force in the MM and hybrid ($\Delta$) region (red dotted line). ${\mu}^{e}$ is the electronic chemical potential of the QM region which should be the same as that of a corresponding full QM calculation ${\mu}^{e}_{ref}$. B.C. stays for Boundary Conditions.}
 \label{cartoon}
\end{figure}

{I will restrict the discussion to systems consisting of solvated molecules in water, as this is the example that is more general and more appropriate to the coupling details reported below. For other systems there may be the need of specific modifications of the general set up. Thus, having in mind water, in order to have some structural consistency between the QM molecules and the classical atomistic molecules, we will assume that the classical atomistic molecular model is not rigid but has flexible bonds and angles, as for example in \cite{paesanimod}. Moreover, in first approximation, I will consider the MM region and the hybrid MM-coarse-grained region as a transition region/artificial filter between the QM region and the large coarse-grained region/reservoir. This first approximation takes care only of the basic physical consistency between the QM region and the classical reservoir (i.e. the macroscopic and electronic chemical potentials of the QM region corresponds to that of an equivalent full quantum macroscopic system of reference). The approximations at this first level may imply some high computational costs in the equilibration of the system, due to the abrupt interface. In the next sections I will then discuss how this problem can be minimized and how the MM region can be treated as a region with a physical meaning on itself and not only as an artificial filter between the QM and the coarse-grained region.}
Given the premise above, the basic idea of extending GC-AdResS \cite{prx,njp,jctchan} to include electrons is based on following main points:
\begin{itemize}
\item The system is divided in four region, QM, MM, Hybrid, CG (see
  Fig.\ref{cartoon}). The MM and hybrid part are an interface between the QM
  and the CG part which allows molecules to gain or loose details and thus to
  be dynamically inserted in/removed from the QM region maintaining the
  thermodynamic equilibrium of a typical Grand Canonical system assured by the
  large CG reservoir.
\item The interface between the MM and the CG is the same as that of
  GC-AdResS, thus the exchange rate of molecules is controlled by the
  (macroscopic) chemical potential of the system. In GC-AdResS, this is achieved by imposing that in the hybrid region the molecular density is the same of the reference density (i.e. addition of the thermodynamic force).  
\item The same molecular exchange rate must occur also between the QM and the MM region.
Since this effect is macroscopic, for similarity to the hybrid region, a
thermodynamic force should be added to the molecules in the MM region so that
in the MM region the molecular density is the same as the one of reference
(the desired density). This will automatically stabilize the molecular density
in the QM region and indirectly equalize the (macroscopic)
chemical potential of the QM and MM region. In fact, as discussed before, the nuclei in the QM region
are classical objects as the atoms in the atomistic region; thus the
thermodynamic force in the MM region has the same role as that of coupling
different molecular representations as proven in \cite{jcpsimon,jctchan}. 
The time-averaged molecular density in the QM region should fulfill the condition of being at the value of reference; this is used as a criterion of control of validity of the hypothesis done and it is linked to the accuracy of the thermodynamic force. In other words, this is a necessary condition for thermodynamic consistency. In addition one should monitor that the exchange of molecules  from one region to another happens in a proper way (i.e. there is the expected flux in both directions) and that the probability distribution of molecules in the quantum region, $P(N)$, follows the expected Gaussian behaviour. This is routinely done in AdResS while the simulation runs (see for example \cite{wat2,annurev,krek}). If both criteria are fulfilled, automatically we have exchange of molecules at the correct chemical potential.
\item If the conditions of scheme outlined so far are fulfilled, as a consequence, we have that the QM region is a Grand Canonical system whose
  molecules are dynamically introduced from or moved into a reservoir (MM+Hybrid+GC) at
  each time step; the classical reservoir is at the same macroscopic thermodynamic conditions ($\mu$, T) of the QM region. 
As in standard AIMD at each time step one has a nuclei configuration in the QM region and a
classical configuration in the rest of the system. The electronic Hamiltonian
corresponds to a Hamiltonian with the instantaneous nuclei positions and instantaneous positions of  classical molecules (for the external interactions); this Hamiltonian is employed to derive
$\rho_{e}({\bf r})$ at given $\mu_{e}$ and in turn the Hellmann-Feynman forces on
the nuclei are calculated from $\rho_{e}({\bf r})$. Next, having the forces,
nuclei and classical molecules are moved to a new configuration according to
Newtons' equations of motion. Some nuclei (quantum molecules) will cross the
boundary of the QM region and for the calculation of the next time step will
become atoms of a classical atomistic molecules. At the same time some classical
molecules will cross the border and enter into the QM region and become nuclei
(of a quantum molecule) for the calculation of next time step. Differently from the
standard classical GC-AdResS where there is a region of artificial hybrid resolution, the
border QM/atomistic is sharp in first instance (later on I will discuss
possible smoothing strategies). In fact a flexible
molecular model can be already considered, at basic level, a sort of hybrid between a fully flexible QM
molecule and a spherical simplified coarse-grained molecule. The explicit details of the coupling are given in the next section.
\end{itemize}
\section{MD Coupling}
Differently from Monte Carlo Grand Canonical simulations \cite{mcgc} where the
insertion and removal of one molecule per time requires the detailed sampling
of the phase space in order to optimally allocate a new molecule or create a
vacancy, the particles' exchange suggested here follows a dynamical evolution of the
system where a certain number of molecules crosses the borders (in all directions)
in full consistency with the statistical mechanics and thermodynamics of a
corresponding full atomistic system. The dynamical coupling of GC-AdResS has
been shown to have several advantages not only for the calculation of static
properties but also for dynamic properties \cite{njp,lujcppi,lucpc} a detailed
discussion of advantages and limitations of the method can be found in \cite{physrep}. Let us consider the dynamical
coupling, that is coupling through forces and let us see how forces are calculated at each step in the scheme proposed. We have the electronic structure of the QM part, calculated at the given electronic chemical potential $\mu^{e}_{ref}$, in a nuclei configuration consistent with the configurations of the atoms of the MM part and of the configuration of the coarse-grained part.
The forces on the molecules of each region are then reported below.
\subsection{QM Region}
The force on the $i$-th atom in the QM region writes:
\begin{equation}
{\bf F}_{i}^{QM}={\bf F}^{HF}_{i}(\rho_{e})+\sum_{j\in QM}{\bf F}^{nuclei}_{i,j}(Z^{a}_{i},Z^{b}_{j})+\sum_{l\in MM+\Delta}{\bf F}^{Coulomb}_{i,l}(Z^{a}_{i}, q^{d}_{l})
\label{fqm}
\end{equation}
${\bf F}^{HF}$ is the Hellmann-Feynman force due to the electron density, ${\bf F}^{nuclei}_{i,j}(Z^{a}_{i},Z^{b}_{j})$ the Coulomb force due to the interaction of nucleus $a$ with charge $Z_{a}$ in molecule $i$ with nucleus $b$ of charge $Z_{b}$ of molecule $j$, ${\bf F}^{Coulomb}_{i,l}(Z_{i}, q_{l})$ is the Coulomb force due to the interaction of nucleus $a$ with charge $Z_{a}$ in molecule $i$ with the charge $q$ of atom $d$ in molecule $l$ in the MM region and hybrid region.
\subsection{MM region}
The coupling forces acting on the $i$-th atom of the $\alpha$ molecule write:
\begin{equation}
 {\bf F}_{i_{\alpha}}^{MM}=\sum_{j\in QM}{\bf F}^{Coulomb}_{i_{\alpha},j}(Z^{a}_{j}, q^{d}_{i_{\alpha}})+{\bf F}_{i_{\alpha}}(q^{d}_{i_{\alpha}},\rho_{e})+\sum_{j_{\beta}\in MM+\Delta}{\bf F}^{MM}+{\bf F}^{th}_{i_{\alpha}}(\rho_{macro})
\label{fmm}
\end{equation}
$\sum_{j\in QM}{\bf F}^{Coulomb}_{i_{\alpha},j}(Z^{a}_{j}, q^{d}_{i_{\alpha}})$ is the force due
to the Coulomb interaction with the charges of the nuclei in the QM region,
${\bf F}_{i_{\alpha}}(q^{d}_{i_{\alpha}},\rho_{e})$ is the Coulomb force between the electron charge of QM and the atomic charges in MM, $\sum_{j_{\beta}\in MM+\Delta}{\bf F}^{MM}$
is the MM force corresponding to the classical force field chosen (i.e. the force between the $i$-th atoms of the $\alpha$ molecule and the $j$-th atom of the $\beta$ molecule, with $\alpha\neq\beta$, $\frac{m_{i_{\alpha}}}{M_{\alpha}}{\bf
  F}^{th}_{\alpha}(\rho_{macro})$ is the thermodynamic force acting on the center of
mass of the molecule $\alpha$, distributed among its atoms according to their masses. Once again, such a force fulfills the condition for the molecular number density: $<\rho_{mol}>=\rho_{macro}=\frac{N_{mol}}{V}$, that is the average molecular
density $\rho_{mol}$ is equal to the molecular density of the macroscopic thermodynamic condition chosen. As reported before, this force is a necessary condition to assure
that two systems with different molecular representation have an exchange of
molecule at the proper macroscopic chemical potential $\mu^{M}$. The technical
calculation of ${\bf F}^{th}$ have been reported in the previous section.
It must be noticed that differently from the standard AdResS, in the hybrid region we have now two additional terms: $\sum_{j\in QM}{\bf F}^{Coulomb}_{i_{\alpha},j}(Z^{a}_{j}, q^{d}_{i_{\alpha}})+{\bf F}_{i_{\alpha}}(q^{d}_{i_{\alpha}},\rho_{e})$, due to the interaction with the quantum part. One may thing of smoothing them down as done for the other term, however for consistency with the corresponding term present in the QM region, one can in first (numerical) approximation leave it as it is. The $\Delta$ region is far enough so that the dominant term in the force is the third one and thus the first two terms do not bring particular numerical problems to the MM/CG transition. Once again, for numerical problem is meant the convergence of the thermodynamic force.
\subsection{$\Delta$ Region}
The force on atom $i$ of molecule $\alpha$ writes:
\begin{eqnarray}
  {\bf F}_{i_{\alpha}}^{\Delta}=\sum_{j\in QM}{\bf F}^{Coulomb}_{i_{\alpha},j}(Z^{a}_{j}, q^{d}_{i_{\alpha}})+{\bf F}_{i_{\alpha}}(q^{d}_{i_{\alpha}},\rho_{e})+
 \label{fdelta1} \\\nonumber\sum_{\beta\in\Delta}\left[w(X_{\alpha})w(X_{\beta}){\bf F}_{i,j_{\beta}}^{MM}+\frac{m_{i_{\alpha}}}{M_{\alpha}}\left(1-w(X_{\alpha})w(X_{\beta})\right){\bf F}^{CG}_{\alpha,\beta}\right]+\frac{m_{i_{\alpha}}}{M_{\alpha}}{\bf F}^{th}_{\alpha}(\rho_{macro})
\end{eqnarray}
$w(X_{\alpha})$ is the weighting function as defined, for example in AdResS, and $X_{\alpha}$ is the center of mass of molecules $\alpha$, ${\bf F}^{CG}_{\alpha,\beta}$ is the force derived coarse-grained potential which acts only between the centers of mass of molecule $\alpha$ and molecule $\beta$ and is distributed on the atoms weighting the force by their corresponding mass.
\subsection{CG region:}
The interaction (among the centers of mass) in the CG region for molecule $\alpha$ writes:
\begin{equation}
{\bf F}_{\alpha}^{CG}=\sum_{\beta \in \Delta+CG}{\bf F}^{CG}_{\alpha,\beta}
\label{fcg}
\end{equation}
the size of the $\Delta$ region in AdResS is always larger than the range of interaction of the CG potential.\\
{It must be underlined that the thermodynamic force appearing in Eq.\ref{fmm} and Eq.\ref{fdelta1} is the same, however such a force is position-dependent thus it acts differently in the MM and in the $\Delta$ region. Moreover such a force, for consistency, must be calculated, in iterative manner, in all regions simultaneously by running a QM+MM+$\Delta$+CG simulation. It may be possible to separate the calculation in the MM and in the $\Delta$ region, this is discussed later on.} 
{QM, MM and $\Delta$ region are subject to a reaction field (as in standard AdResS), which mimics the dielectric properties of the coarse-grained region (where charges are not explicitly considered). Such an approximation has been successfully tested in AdResS for a large variety of systems. In particular it turned out to be very accurate for ionic liquids for which electrostatics plays the major role and its description is highly challenging \cite{krek}. As a consequence, also the first two terms of Eq.\ref{fdelta1}, which correspond to forces between the QM region and molecules in the $\Delta$ region are subject to the same cutoff of the reaction field.}
\section{Smoothing the transition at the QM/MM Interface}
\subsection{Excess charge correction}
 It must be underlined that, from a practical point of view, the treatment of
 the QM region at constant $\mu_{e}$ implies that one cannot control $N_{e}$
 and electrons and fractions of electrons may be delocalized outside the QM region and thus not treated
 explicitly. This situation can lead to the case where $\int_{V_{QM}}\rho_{e}({\bf
   r})d{\bf r}-\sum_{I}\sum_{i}{Z_{I}^{i}}\neq 0; \forall I,i \in QM$, where
 $Z_{I}^{i}$ is the charge of the $i-$th nucleus of the $I$-th molecule. The
 excess of positive or negative charge in the QM region should be carefully treated.
The missing charge or excess electronic charge corresponds to delocalized
electrons (or fraction of electrons) and is not
treated explicitly as a QM quantity, but is somehow distributed as a
classical charge in the MM part. In the simulation, the excess of charge,
either positive or negative, is given at each step by: $\int_{V_{QM}}\rho_{e}({\bf
  r})d{\bf r}-\sum_{I}\sum_{i}{Z_{I}^{i}}\neq 0; \forall I,i \in QM$, an
effective way to take into account it, is via a redistribution of the charge by
rescaling accordingly the charges on the MM atoms at each step.
Moreover since the probability to find the delocalized electron is a function
of the distance from the QM region, one could set up a (cutoff) transition region between the the QM and the MM part where the extra charge on atoms appearing in the expression of the Coulomb force is automatically redistributed at each time step. For example defining a transition shell where the excess of charge is (in first instance) uniformly distributed on each atom of the region. The interface with the MM region will not create a problem, in fact interfacing models with different charges has been already successfully tested in GC-AdResS for ionic liquids \cite{krek}. One may even think of building several transition shells where the excess charge is progressively decreased as function of the distance from the QM region down to zero
outside this hybrid QM/MM region. As in standard GC-AdResS the molecules in
such an hybrid region(s) would be a sort of hybrid between he QM molecules and
the MM molecules. The relevant point is that statistical and thermodynamic consistency will still be assured by the thermodynamic force for balancing the
macroscopic chemical potential between the QM and MM region. Now this force will be applied only in the QM/MM hybrid region and not anymore in the full MM region. A more specific discussion about an additional transition region will be reported later on. Of course a realistic distribution of the excess charge will lead to a smoother numerical derivation of the thermodynamic force, instead an excessively unrealistic distribution will lead to serious numerical problems in the convergence of the thermodynamic force. This situation is very similar to that of the transition region between the atomistic and coarse-grained region: a bad coarse-grained model of a molecule implies problems of convergence of the thermodynamic force. A further smoothing can be achieved considering the intramolecular interactions in the MM region,as explained below.
 \subsection{Intramolecular interaction in the MM region}
If we consider a classical flexible model of, for example, water molecules, the
classical molecule will have some intramolecular interactions, mostly in the
form of a quadratic potential \cite{flexwat}. The exploration of the intramolecular phase
space of a classical molecule will anyway be different from that of a quantum
molecule which does not have predefined bonds. As a consequence a molecule
from the QM region which enters in the MM region may have intramolecular
distances and angles that are outside the accessible space of the MM
molecules.
Thus an abrupt passage from the QM to the MM region may create some local
disequilibrium at the QM/MM border. The thermodynamic force will
take care of reestablishing in average the correct equilibrium, but it may turn
out that due to this abrupt passage the calculation of the thermodynamic force
becomes tedious. For this reason, as for the excess of charge in the previous
section, one can consider a transition region between the QM and the MM region
where the force due to the classical intramolecular potential is weighted by a function that
smoothly relaxes the intramolecular forces as a molecules goes from the
MM to the QM region, and vice versa, smoothly increases the intramolecular
forces as a QM molecules goes from the QM region into the MM region. Further discussion is reported in the section below.
\subsection{MM beyond the role of artificial filter: Multiscale approach}
\begin{figure}
   \centering
   \includegraphics[width=0.75\textwidth]{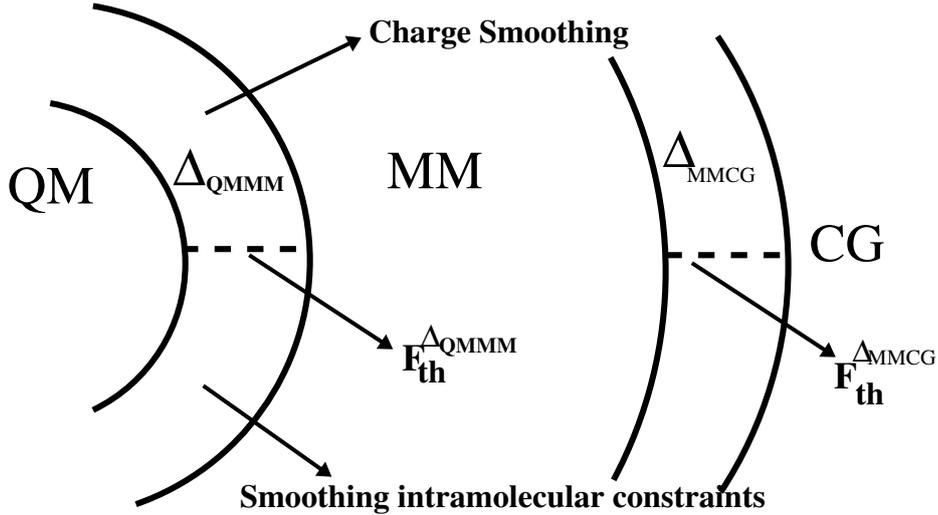}
   \caption{{Schematic representation of the additional interface acting as a filter between the QM and (a physically meaningful) MM region. The charge smoothing could be accomplished by, for example, assigning (in a uniform manner) 50\% of the excess charge to atoms of molecules within a certain distance $d1$ from the QM region, 30\% to atoms of molecules within a region defined by $d2-d1$ in $\Delta_{QMMM}$, with $d2>d1$, and so on. The Smoothing of the intramolecular potential can be done, for example, by softening the potential in such a way that at the border between the MM and the $\Delta_{QMMM}$ region the weighted potential has as a limit the intramolecular force field of the MM model and as the molecules moves towards the QM region, the weighted potential is such that internal angles and bonds can assume values beyond those allowed by the MM force field but still within the range of values possible in molecules of the QM region. More complex schemes that optimize the computational convenience may be developed also, however in first instance the sugestions above may be sufficient for a smooth calculation of the thermodynamic force in the $\Delta_{QMMM}$ region. Finally the thermodynamics force now acts sepaerately in the $\Delta_{QMMM}$ and in the $\Delta_{MMcg}$ regions.}}
 \label{delta2}
 \end{figure}

{As discussed earlier on, the abrupt transition between the QM and MM implies that only the physics of the QM and of the coarse-grained region is well defined, while the MM region together with the $\Delta$ region plays the role of an artificial filter which allows the proper thermodynamic equilibrium and exchange between the QM and the CG region. However, in a multiscale spirit, the MM region can be physically meaningful and its interface with the QM region can be made computationally smoother by introducing a second $\Delta$ region between the QM and the MM region. Such a region would be a portion of the MM space confining with the QM region, let us term it $\Delta_{QMMM}$. A pictorial representation is reported in Fig.\ref{delta2}.}
{In such a case, the thermodynamic force does not apply on the whole MM region, but only in the $\Delta_{QMMM}$ and it can be calculated independently from the thermodynamic force of the MM-CG transition region, since the MM region would be large enough that instantaneous fluctuations of density in the other regions would not influence each other. This transition region can be built according to the principles reported in the two previous subsections, that is introducing a smoothing, space-dependent, distribution of the excess of charge on classical atoms in $\Delta_{QMMM}$ and  a space-dependent smoothing of the intramolecular potential of the classical molecules. In such a way, this additional interface acts as a region of ``preparation'' of the classical molecules to provide, as much as possible, the proper nuclei skeleton for the quantum molecule once the molecule enters in the QM region (and vice versa from QM to MM with a classical ``de-quantization'' process). Here for proper is meant that the nuclear structure provided by the classical molecule, once entered in the QM region, is compatible with that of a hypothetical quantum molecules as in a full quantum system of reference. The advantage of the interface of ``preparation'' consists in being only at classical level. This fact avoids nonphysical hybrid quantum/classical definition of a molecule which can have only an empirical justification in the calculation of a quantum property, for example the empirical average done in QM/MM methods. Instead in our case the QM region is well described within a physically meaningful model of Grand Canonical ensemble, that is the electronic properties have an objective physical meaning. In addition, any possible QM/MM technique applied to the MM molecules that smooths the interface could be integrated into the $\Delta_{QMMM}$ region \cite{tru}.}
 
\section{Discussion and Conclusion}
I have proposed a theoretical framework for a computational protocol to couple an open quantum system of molecules with a classical (atomistic and large coarse-grained) environment. Molecules in the QM region are composed of electrons and (classical) nuclei and can move into the classical region (and vice versa). The exchange of molecules occurs under principles based on physical consistency at macroscopic level and at (microscopic) electronic level. In fact the QM subsystem would be treated at a constant electronic chemical potential corresponding to that of a full quantum system, while the macroscopic chemical potential of the liquid is kept at the desired value at the given thermodynamic conditions. The coupling principles and the strict conditions of control of the approximations proposed here should actually be fulfilled also by well established open QM/MM approaches, thus in this sense the criteria proposed here are in fact general whenever one has a QM system embedded in a classical environment/reservoir. If such criteria are fulfilled then the validation of the simulation results does not require anymore the direct comparison (for each simulation) with equivalent full quantum simulations as instead it is currently done \cite{revqmmm}. In this perspective, the intention of this paper is to stimulate the cooperation between the very different fields of electronic structure calculations of open systems, usually located in the field of electrochemistry and voltage application, and coarse-grained simulations, interested instead in mesoscopic properties. This basic protocol offers the possibility of finding a common ground and may stimulate the possibly to improve the theoretical and technical principles of coupling put forward in this work for the finalization of a working code for actual simulations.


\vspace{6pt} 

\acknowledgments{This work was supported by the Deutsche Forschungsgemeinschaft (DFG) with the grant CRC 1114 (project C01).}

\end{document}